\documentclass[letterpaper,twocolumn,10pt]{article}
\usepackage{usenix}

\usepackage{tikz}
\usepackage{amsmath}

\usepackage[ruled,linesnumbered]{algorithm2e}

\usepackage{graphicx}
\usepackage{booktabs}
\usepackage{listings}
\usepackage{multirow}
    \usepackage[table]{xcolor}

\usepackage{adjustbox}
\usepackage{tcolorbox}
\usepackage{hyperref}
\usepackage{enumitem}
\usepackage{tabularx}
\usepackage{makecell} % For better multi-line cells

\usepackage{listings}
\usepackage{xcolor}

\usepackage{xcolor}

\lstdefinelanguage{AttackDesc}{
  morecomment=[l]{\%}
}

\lstdefinestyle{attackdesc}{
  language=AttackDesc,
  basicstyle=\ttfamily\footnotesize,
  commentstyle=\color{blue!70}\itshape,
  breaklines=true,
  breakatwhitespace=false,
  columns=fullflexible,
  keepspaces=true,
  frame=single,
  showstringspaces=false,
  captionpos=b,
}
\usepackage{pgfplots}
\pgfplotsset{compat=1.18}
\begin{document}
%-------------------------------------------------------------------------------

%don't want date printed
\date{}

\title{From Description to Detection: LLM based Extendable O-RAN Compliant Blind DoS Detection in 5G and Beyond}

%for single author (just remove % characters)
\author{
    Thusitha Dayaratne$^{1}$\thanks{Corresponding author: \texttt{thusitha.dayaratne@monash.edu}},
    Ngoc Duy Pham$^{1}$,
    Viet Vo$^{2}$,
    Shangqi Lai$^{3}$,
    Sharif Abuadbba$^{3}$,\\
    Hajime Suzuki$^{3}$,
    Xingliang Yuan$^{4}$,
    Carsten Rudolph$^{1}$ \\
    $^{1}$Monash University \\
    $^{2}$Swinburne University of Technology \\
    $^{3}$CSIRO's Data61 \\
    $^{4}$The University of Melbourne}

\maketitle

%-------------------------------------------------------------------------------
\begin{abstract}
%-------------------------------------------------------------------------------
The quality and experience of mobile communication have significantly improved with the introduction of 5G, and these improvements are expected to continue beyond the 5G era. However, vulnerabilities in control-plane protocols, such as Radio Resource Control (RRC) and Non-Access Stratum (NAS), pose significant security threats, such as Blind Denial of Service (DoS) attacks. Despite the availability of existing anomaly detection methods that leverage rule-based systems or traditional machine learning methods, these methods have several limitations, including the need for extensive training data, predefined rules, and limited explainability.
Addressing these challenges, we propose a novel anomaly detection framework that leverages the capabilities of Large Language Models (LLMs) in zero-shot mode with unordered data and short natural language attack descriptions within the Open Radio Access Network (O-RAN) architecture. 
We analyse robustness to prompt variation, demonstrate the practicality of automating the attack descriptions and show that detection quality relies on the semantic completeness of the description rather than its phrasing or length. We utilise an RRC/NAS dataset to evaluate the solution and provide an extensive comparison of open-source and proprietary LLM implementations to demonstrate superior performance in attack detection. We further validate the practicality of our framework within O-RAN’s real-time constraints, illustrating its potential for detecting other Layer-3 attacks.
\end{abstract}

\section{Introduction}
The fifth-generation (5G) cellular networks provide foundations for technical advancements in various sectors, including communication, healthcare, agriculture, and education, through enhanced Mobile Broadband (eMBB), massive Machine-Type Communications (mMTC), and Ultra-Reliable Low-Latency Communication (uRLLC). 
Alongside these public deployments, private 5G networks, which are customised for enterprise, industrial, and critical-infrastructure environments, are also emerging with industry leaders such as Tesla and Porsche already moving towards adoption~\cite{5gbox}. Despite significant technological advancements and numerous benefits of the latest cellular networks (both public and private), the inherent distributed nature of these networks has significantly increased the attack surface and the complexity of the security landscape. Even small local attacks can have a significant impact and the risk of these occurring has increased due to this expanded attack surface, the diversity of connected devices, and the critical nature of services relying on 5G networks~\cite{vargas2023impacts,khan2019survey}.

Several vulnerabilities~\cite{5GReasoner,5gsepctor,adaptover} exist in Layer-3 protocols such as the Radio Resource Control (RRC)~\cite{ts38331} protocol, which manages key control-plane functionalities including connection establishment, and radio bearer configuration, and the Non-Access Stratum (NAS)~\cite{ts24501} protocol that handles mobility management and session management between User Equipments (UEs) and core networks. These protocols are increasingly attractive to adversaries given that RRC and NAS protocols orchestrate the communication states of UEs. This risk is further increased due to the lack of robust integrity protection and encryption for many RRC/NAS messages~\cite{5gsepctor,detran}, which makes these messages susceptible to tampering and eavesdropping. Despite the security measures provided by 3GPP specifications, these protocols are often arduous to update, and significant revisions may not be fully realised until the development of 6G standards due to backward incompatibilities~\cite{wang2021privacy}. Furthermore, some of potential weaknesses are inherent and are difficult to fix on this layer without losing openness and functionality. Hence, attackers who exploit these vulnerabilities can perform multiple attacks, such as Blind Denial of Service (DoS), Null cipher, Downlink DoS, and Lullaby attacks, which force session terminations, battery drainage, or even subvert the establishment of cipher and integrity protection~\cite{5GReasoner,touchinguntouchables,adaptover}. Consequently, it is essential to detect these attacks before they make an impact. However, for third parties it is difficult to minimise the impact of these attacks, as closed architectures restrict controlling the RRC/NAS message flows in traditional, monolithic cellular architectures%, is restrictedwhich are less customisable and bounded by vendor-specific restrictions. 

In recent years, the Open Radio Access Network (O-RAN) architecture is emerging to move from a closed and proprietary radio access network to an open version with higher interoperability and extendability. it also enables more stakeholders to enhance control and security and visibility of the radio access layer of cellular networks. As of August 2025, there are 18 commercial, 11 pre-commercial, and 31 field-trial O-RAN deployments throughout the globe~\cite{o_ran_map}. The O-RAN architecture decouples RAN components and exposes standardised interfaces that allow network operators and third-party developers to implement custom, real-time monitoring and response functions. This decoupling of hardware and software allows the development of flexible, highly scalable, and efficient solutions that can address a wide range of use cases, from standard Quality-of-Service (QoS) requirements to more complex scenarios, such as anomaly detection and network slicing in a vendor-agnostic manner. In particular, the O-RAN architecture allows deep inspection and controllability of control-plane messages by enabling the integration of Artificial Intelligence and Machine Learning (AI/ML) modules through RAN Intelligent Controllers (RICs). Some recent works have integrated both rule-based~\cite{5gsepctor} and traditional ML-based anomaly detection approaches~\cite{detran,6gxsec} with the O-RAN to detect potential Layer-3 attacks. 

Nevertheless, both rule-based and traditional ML-based anomaly detection approaches have their shortcomings, including the need to explicitly define rules, the requirement to keep track of every session, the need for substantial amounts of data to train models, and the necessity of pre-training. Additionally, these approaches lack explainability. Furthermore, a recent work demonstrates \cite{hypoglyph2025}, traditional ML models crash when an adversary injects Unicode alterations (hypoglyphs) in control-plane messages into the O-RAN Shared Data Layer (SDL). These limitations are critical in practical 5G deployments, where devices connect simultaneously, leaving detectors with only unordered set of messages. As a result, more nuanced approaches are essential. 

Recently, Large Language Models (LLMs) have emerged as an alternative to explore new approaches to anomaly detection. The zero-shot classification capability enables LLMs to classify messages while producing natural-language explanations. Additionally, LLMs remain robust when the NAS/RRC messages stored in the SDL are manipulated with adversarial symbols that hinder traditional ML based approaches~\cite{hypoglyph2025}. In particular, LLM's zero-shot classification capability can address the shortcomings of both rule-based and traditional ML-based anomaly detection. Thus, in this work we introduce an anomaly detection framework that leverage these strengths of LLMs within the O-RAN architecture to detect Blind DoS attacks. To the best of our knowledge, this is the first approach that operates on unordered RRC/NAS data in pure zero-shot mode, eliminating the need for ordering assumptions, manual rules, or iterative feedback. In particular, we select the Blind DoS attack, as recent work has highlighted the need for a real-time solution to address this attack, given that there is still no effective countermeasure or solution available~\cite{detran}.

Our solution encapsulates the unordered RRC/NAS messages along with a short natural-language description of the attack description into a compact prompt. Then we leverage an off-the-shelf LLM in zero-shot mode to decide whether the message set contains the attack and to return a short explanation. We evaluate our approach using an existing RRC/NAS dataset and compare the effectiveness of both open-source and proprietary LLMs. Further, show how we can extend our framework to automate the prompt creation using the power AI agents. We demonstrate the feasibility of leveraging LLMs within the O-RAN architecture, where LLM determines the classification within 30ms, which is well within the timing constraints (10ms - 1s) and the robustness of the approach against evasion attacks. Although our framework is designed to detect Blind DoS attacks, we show that it can be extended to detect other types of Layer-3 attacks. In particular, we make the following contributions.

\begin{enumerate}
    \item We propose a novel LLM-based Blind DoS detection method, addressing the inherent shortcomings in traditional ML and rule-based approaches. Our solution does not require manual session tracking, explicit rule definitions, training phases with labelled datasets. Instead, it leverages natural language attack descriptions to build an effective detection model.
    \item We introduce a zero-shot anomaly detection approach for NAS/RRC messages, which perfectly detect Blind DoS attacks on unordered data. It uses an off-the-shelf LLM relying only on contextual information for the detection.
    \item We integrate an automated attack–description pipeline minimising manual work. Starting from natural language attack descriptions, concise descriptions are generated and the effectiveness of the description is automatically evaluated. 
    \item We provide an extensive empirical evaluation of the proposed LLM-based Blind DoS detection method within a realistic O-RAN setting. Furthermore, we analyse the practical implications, potential limitations, and offer solutions to ensure applicability in real-world deployments including the detection of other Layer-3 attacks.
\end{enumerate}

\section{Background}
\subsection{Cellular Network Architecture, Procedures and Identities}
UE, the 5G base station (gNodeB/gNB), and the 5G core (5GC) network are the three main components of the standard 5G cellular architecture. UEs include mobile phones, IoT devices, and other wireless devices that possess a Universal Subscriber Identity Module (USIM), which uniquely identifies the device. In contrast to previous cellular generations (4G and prior), 5G replaced the traditional IMSI (International Mobile Subscriber Identity) with the SuPI (Subscription Permanent Identifier), which is designed to reduce identity tracking (IMSI catchers~\cite{imsicatcher}) and improve user privacy. The gNodeB serves as the access point for UEs to connect to the core network. The 5GC leverages Service-Based Architecture (SBA) and consists of multiple functional entities, including the Access and Mobility Management Function (AMF), which handles user registration and mobility; the User Plane Function (UPF), which handles data routing, the Session Management Function (SMF), which manages session handling, and the Authentication Server Function (AUSF) for secure authentication.

\subsubsection{Initial Access}
\begin{figure}
    \centering
    \includegraphics[width=.95\linewidth]{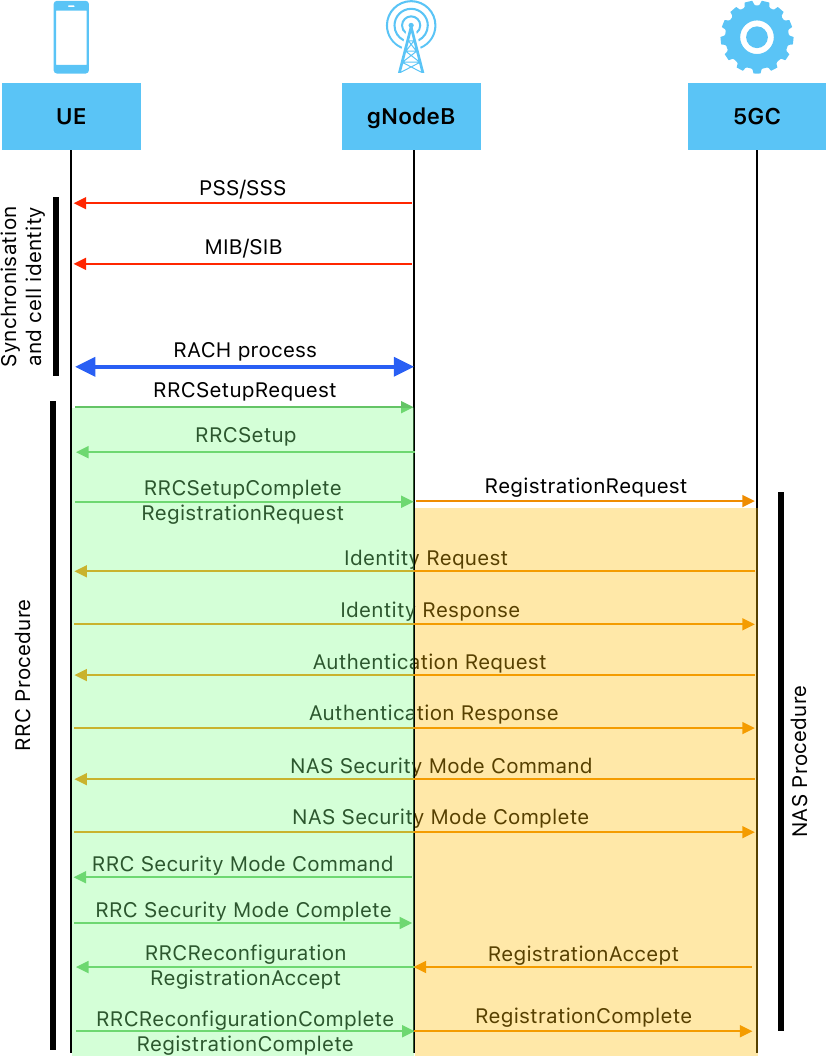}
\caption{High-level initial access procedure in 5G SA}
    \label{fig:initial_access}
\end{figure}

5G initial access is the standard procedure that establishes a connection between a UE and the 5G network. In 5G, two deployment modes exist: Non-Standalone (NSA), which relies on existing 4G infrastructure for control functions while leveraging 5G for data; and Standalone (SA), a fully independent 5G system with a separately designed core network and radio access. In 5G SA, the initial access process begins with the UE scanning for synchronisation signals. More specifically the Primary Synchronisation Signal (PSS) and Secondary Synchronisation Signal (SSS), which are broadcast by the gNodeB. These signals help the UE synchronise timing and frequency with the network and identify the cell’s Physical Cell Identity (PCI). Following synchronisation, the UE decodes the Master Information Block (MIB) and System Information Blocks (SIBs) from the Physical Broadcast Channel (PBCH) to acquire essential configuration details, including cell selection information, cell access information for the serving cell and Public Land Mobile Network (PLMN) identity information. The UE then initiates a random access procedure using the Physical Random Access Channel (PRACH), sending a preamble to the gNodeB, where the gNodeB responds with a timing adjustment. Upon synchronisation, the UE starts the RRC procedure by sending an \textit{RRCSetupRequest} (either with a 5G-S-TMSI or a 39-bit random value), followed by an \textit{RRCSetup} message from the gNodeB. The UE completes RRC connection establishment by sending an \textit{RRCSetupComplete} message and an embedded NAS \textit{RegistrationRequest} to initiate the NAS procedure. The complete flow of the initial access procedure is depicted in Figure \ref{fig:initial_access}.

\subsubsection{Identities in 5G}
Multiple identities are used to ensure secure and efficient user identification, session management, and mobility in 5G SA networks. These identities serve different purposes, ranging from permanent subscriber identification to temporary and context-specific identifiers. The 5G-Globally Unique Temporary Identifier (5G-GUTI), assigned by the AMF to temporarily identify the user during network operations, is one of the key identifiers. Within the 5G-GUTI structure, the 5G-Temporary Mobile Subscriber Identity (5G-TMSI -- we interchangeably use 5G-TMSI and TMSI) is used for mobility management and paging, allowing the network to reach a UE without exposing its permanent identity. The TMSI value is assigned to a device in the NAS \textit{RegistrationAccept} message when a UE is connecting for the very first time. UEs with already assigned TMSIs can use their assigned TMSI values starting from the \textit{RRCSetupRequest}~\cite{ts38331}.
The Radio Network Temporary Identifier (RNTI) is another identity, which is dynamically assigned by the gNodeB to be used in downlink and uplink communications starting from the RACH process.

\subsection{O-RAN Architecture}
Figure \ref{fig:oran_arch} depicts the high-level O-RAN architecture and the interfaces among the different O-RAN components. The O-RAN architecture aims to transform traditional proprietary and monolithic RAN architectures into open and disaggregated networks, empowered by the principles of intelligence and openness. In particular, O-RAN distributes the traditional base station into multiple logical virtualised elements, including the O-CU, O-DU, and O-RU, allowing for a more flexible and vendor-neutral network. 
Further, the O-RAN architecture enables the implementation of network intelligence and automation through the RAN Intelligent Controllers (RICs), to cater to diverse service requirements and improve overall network performance~\cite{polese2023understanding,santos2025managing}. 

RICs serve as the central intelligence layer within the O-RAN architecture, facilitating efficient, flexible, and dynamic network management. 
O-RAN specifies two distinct types of RICs based on their operational timescales and functional objectives: the Near-Real-Time RIC (Near-RT RIC), which allows it to access and process real-time data within a control loop of 10ms - 1s with the support of specialised apps named xApps and the Non-Real-Time RIC (Non-RT RIC) that executes over the 1s control-loop through rApps.

\subsection{LLMs Preliminaries}
LLMs, the most recent evolution of massive deep neural networks, are designed to process, understand, and generate natural language (like humans)~\cite{llm_survey}. In general, LLMs are based on transformer architecture and trained on massive datasets, which allow them to learn complex language patterns, relationships, and contextual meanings. Their capabilities extend beyond traditional natural language processing (NLP) tasks, enabling applications in areas such as automated text generation, sentiment analysis, and even predictive analytics. One of the emerging applications of LLMs is anomaly detection~\cite{sinha2024real}, where LLMs can analyse patterns in data and identify deviations that may indicate fraud, cybersecurity threats, system failures, or other irregular activities. Traditional anomaly detection methods often rely on rule-based systems or statistical models, which may struggle with complex, high-dimensional, or unstructured data and also require a substantial amount of training. On contrary, LLMs can leverage their deep contextual understanding to detect anomalies in various domains, such as financial transactions, network security, and industrial monitoring~\cite{su2024large}.

\subsubsection{Zero-shot}
A Zero-shot approach in the context of LLMs refers to the capability of LLMs to perform tasks without explicitly training on task-specific examples~\cite{kojima2022large}. More specifically, in a Zero-shot approach, the LLM is expected to leverage its extensive pretraining on diverse datasets and general understanding of language to interpret and respond appropriately to prompts it has never encountered before. Relaxation of the dependency on labelled data is the main advantage of this approach. Further, many works have shown that it is feasible to use the Zero-shot approach and enable flexible and scalable applications in various domains~\cite{zhou2023anomalyclip}, including telecommunications~\cite{zeroshottelcom}.

\subsubsection{Chain-of-Thought (CoT)}
The Chain-of-Thought (CoT) approach allows LLMs to explicitly outline intermediate reasoning steps, which enhances their problem-solving capabilities and interpretability~\cite{cot}. Instead of providing immediate answers, the CoT articulates the LLMs thought process, making explicit connections and logical deductions. This step-by-step reasoning approach improves the accuracy of complex reasoning tasks by enabling better identification and correction of errors in deduction/reasoning. Recent work has demonstrated that CoT significantly boosts the performance of LLMs in various tasks, including arithmetic, logical reasoning, and commonsense reasoning.

\section{System Overview}
In this section, we define the threat model, discuss the challenges and insights to overcome them, and present an overview of the solution.

\subsection{Threat Model}
In this work, we focus on the Blind DoS attack~\cite{5GReasoner,touchinguntouchables,detran,milcomdos}, a Layer-3 attack in the 5G context. The Blind DoS attack represents a significant security vulnerability in 5G cellular networks that allows attackers to disconnect legitimate users from the network silently and without their knowledge. This attack leverages the fundamental weakness of lack of integrity protection in certain control messages, such as the \textit{RRCSetupRequest}, \textit{RRCResumeRequest}, and \textit{RRCReestablishRequest}. Adversaries exploit this weakness by impersonating a legitimate UE and spoofing its valid 5G-TMSI, causing a silent disconnection.

In our threat model, we implicitly assume that the core network and gNodeB components are secure, trustworthy, and functioning correctly. We also assume that the adversary cannot compromise or directly manipulate the core network or gNodeB operations. However, adversaries can control their own UEs or employ Software-Defined Radios (SDRs) to customise and inject RRC signalling messages necessary for executing the described attack.

\begin{figure}
    \centering
    \includegraphics[width=1\linewidth]{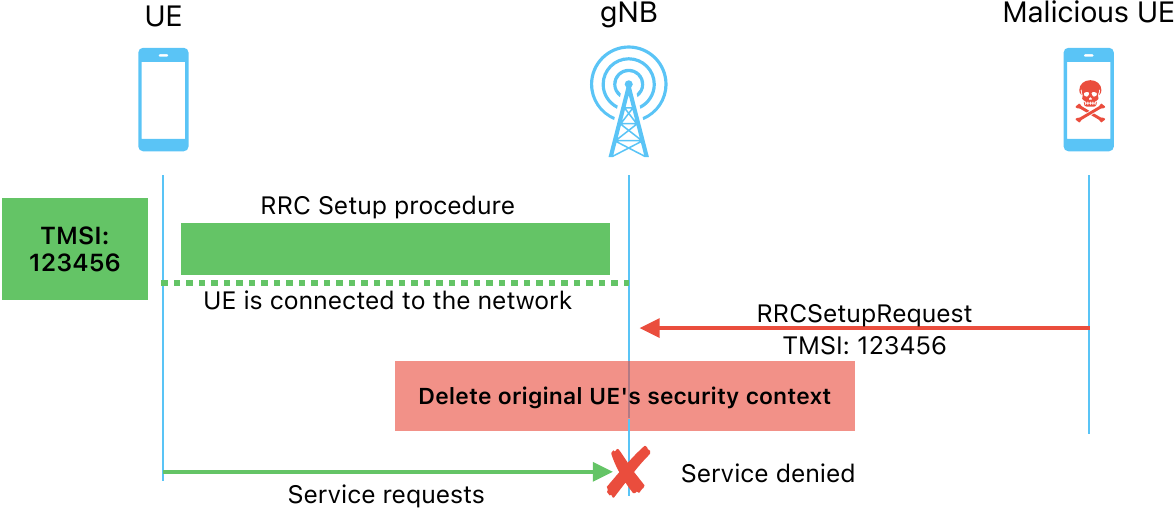}
\caption{High-level overview of the Blind DoS attack}
    \label{fig:bdos}
\end{figure}

The high-level overview of the attack is shown in Figure~\ref{fig:bdos}. The attack steps include identification of the target UE's TMSI and fabrication of an \textit{RRCSetupRequest}. The target UE's TMSI can be obtained leveraging multiple methods, such as deploying a fake base station, performing a silent paging attack targeting a known phone number, or passively eavesdropping on RRC procedures during legitimate signaling interactions~\cite{touchinguntouchables}. After acquiring the TMSI, the adversary constructs and transmits an \textit{RRCSetupRequest} containing the spoofed victim’s 5G-TMSI to the serving base station, to which the victim is already connected. Given the absence of integrity protection at early messages of the initial access procedure, the network accepts the request as legitimate and deletes the existing security context of the legitimate UE and establishes a new connection context with the attacker’s spoofed identity. 

The impact of this attack is severe as the victim UE gets disconnected from the network without explicit notification or alert. The legitimate user experiences denial of service without understanding the cause, and manual re-establishment of connectivity is required for reconnection.  From the perspective of the victim UE, the disconnection appears indistinguishable from a normal network event. Furthermore, given sufficient resources and malicious intent, adversaries can repeatedly execute this attack, sustaining a persistent DoS state for the victim UE.

Despite the fact that this attack was initially discovered in 2019~\cite{5GReasoner} (demonstrated in LTE contexts at the same time~\cite{touchinguntouchables}), this vulnerability still persists~\cite{milcomdos} in current deployments due to the complexity of cellular network protocols and the specification development process. In particular, implementing complete integrity protection for all control-plane messages increases signalling overhead and latency. Further, the need to maintain backward compatibility with existing infrastructure limits security enhancements at the protocol level. Additionally, commercial 5G networks often implement low 5G-TMSI refresh rates~\cite{tmsirefresh}, which enables the spoofing of identities by malicious users. Moreover, the recently introduced Short Data Transmission (SDT) procedure that allows small and infrequent data transmission while UEs are in the \textit{RRC\_INACTIVE} state can also be exploited via the \textit{RRCResumeRequest} message, which broadens the potential for the attack~\cite{detran}.

\subsection{Challenges and Insights}
\textbf{C1: Unavailability of Telemetry Data}
A significant challenge in analysing and mitigating the Blind DoS attack in 5G networks using O-RAN is the unavailability of detailed security telemetry data, specifically the correlation between TMSI and RNTI values within RRC and NAS messages. While the RNTI identifier appears consistently across RRC messages, the TMSI is available only in certain messages, such as \textit{RRCSetupRequest} (for already connected UEs) and NAS \textit{RegistrationAccept}, complicating efforts to precisely track and correlate UE identities throughout message exchanges. Furthermore, the default O-RAN standards and service models do not inherently support fine-grained, security-aware statistical monitoring of the RAN data plane, which hinders the association of individual messages to corresponding UE identities. To overcome this limitation, we leveraged MobiFlow~\cite{5gsepctor}, which employs the E2SM-KPM (v2.0) service model within O-RAN to systematically collect detailed telemetry information. This data is subsequently stored and maintained in the SDL database provided by the RIC infrastructure, enabling accurate correlation between TMSI and RNTI identifiers and enhancing our ability to detect Blind DoS attacks in 5G networks.

\textbf{C2: Optimal Window Size}
Another significant challenge in detecting Blind DoS attacks arises from the lack of a clearly defined window size to analyse past messages for proper contextual understanding. Determining whether a TMSI is legitimately used requires knowing if a successful RRC connection was previously established. Despite intuition suggesting that analysing a single past message (such as the initial \textit{RRCSetupRequest}) would suffice, more contextual information theoretically offers better detection accuracy. However, our experiments using LLMs revealed that while a single past message provided perfect detection accuracy, feeding multiple past messages caused confusion and led to severely degraded performance. Explicitly annotating message sequences (clearly indicating the latest and previous messages) provided limited improvement.

\textbf{C3: Lack of Datasets}
The absence of a publicly available real-world dataset that reflects actual RRC and NAS message exchanges is another challenge. Despite 5G technology being commercially introduced in late 2019, widespread deployment of 5G-SA networks is still scarce. Hence, there is no publicly accessible real-world dataset, which captures realistic attack scenarios. Researchers leverage simulations or testbeds to overcome this issue. Following the same approach, we developed a local test setup using OpenAirInterface (customised for attack execution) deployed within Docker containers, to generate datasets that mimic realistic RRC-NAS message exchanges under normal and attack conditions. 

\textbf{C4: Telecommunication Data in LLM Context}
Another challenge encountered was translating network-specific data into a format suitable for LLMs. Although LLMs perform well in general pattern recognition tasks, they inherently lack domain-specific contextual knowledge about telecommunications processes, especially regarding specialised attacks like Blind DoS. Further, the absence of a sufficiently large dataset prevents fine-tuning an LLM directly for classifying Blind DoS attacks. Hence,  we adopted a prompt-based approach, transforming raw network data into natural language statements, clearly annotating messages (e.g., "\textit{RRCSetupRequest} with RNTI: 123456, and TMSI: 987654"). Additionally, a concise yet informative description of the Blind DoS attack was integrated into the prompt to teach the LLM to recognise the attack pattern without explicit fine-tuning, which helped us to overcome the limited dataset availability.

\subsection{Framework}
\begin{figure*}
    \centering
    \includegraphics[width=1\linewidth]{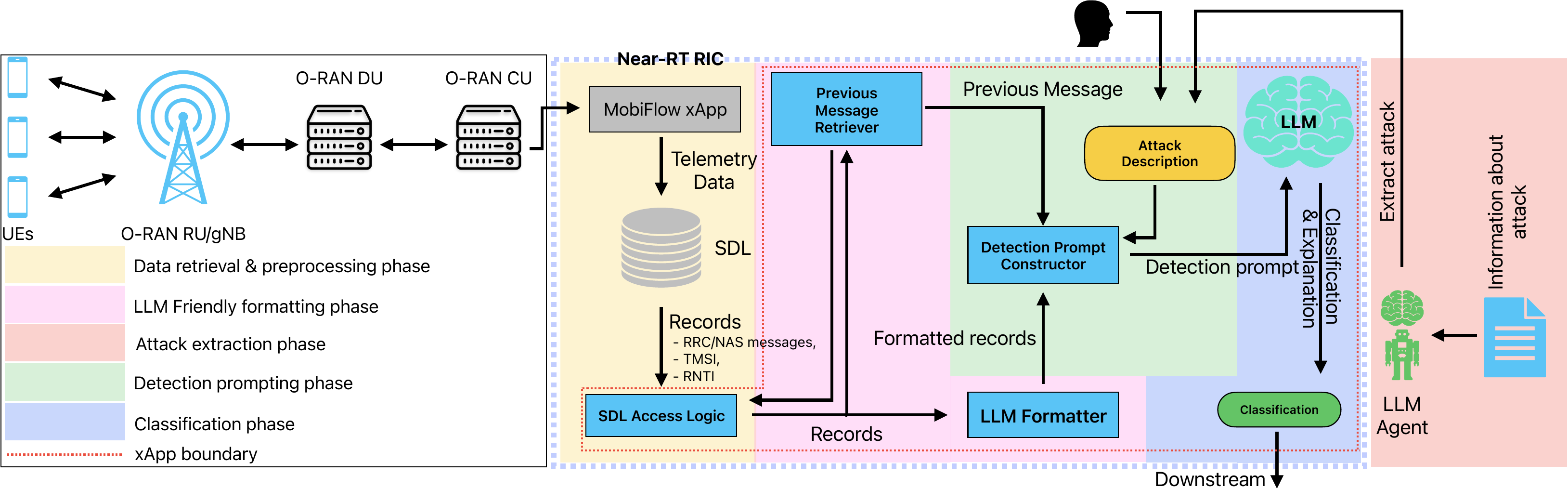}
\caption{High-level overview of the framework}
    \label{fig:framework}
\end{figure*}

A high-level overview of the proposed framework is depicted in Figure \ref{fig:framework}. The framework resides in the Near-RT RIC as an xApp and relies on the MobiFlow xApp to collect RRC/NAS messages along with associated TMSI and RNTI identifiers from the O-RAN context. This captured data is stored in an SDL database~\cite{5gsepctor}. The SDL Access Logic module polls these records for Blind DoS detection. Upon retrieval, these messages are preprocessed. Preprocessing includes the Previous Message Retriever, which is responsible for retrieving the last corresponding message (with same TMSI), and the LLM Formatter, which translates messages into a LLM-friendly format optimised for consumption by LLMs.
Attack descriptions are obtained in an attack extraction phase and may be authored manually by analysts or automatically extracted and summarised by an LLM agent from available attack information.

The processed data, the last corresponding message, and attack descriptions are then integrated by the Detection Prompt Constructor to build the detection prompt. This prompt is the input to an LLM, which executes the classification of messages, providing both a categorisation decision (Normal vs Anomalous) and an accompanying explanation.

\section{Detailed Design}
In this section, we provide a detailed description of each part of the framework, including the design choices and the rationale behind them. For ease of explanation, we divide the framework into five phases, as illustrated in Figure \ref{fig:framework} (shown in different colours). We omit the details of the service model and the collection of telemetry data, assuming that the telemetry data is already available in the SDL. The following sections will outline the detailed process thereafter.

\subsection{Phase 1: Data Retrieval and Preprocessing}
The SDL Access Logic module periodically polls messages stored in the SDL to retrieve the RRC/NAS messages along with their associated TMSI and RNTI values. These three features were specifically selected following recent work~\cite{6gxsec}, which has effectively leveraged traditional ML methods, such as AutoEncoders and Long Short-Term Memory (LSTM) networks, for anomaly detection purposes.

During preprocessing, we observed that the current OpenAirInterface framework employs somewhat similar names for certain NAS and RRC messages, such as \textit{Securitymodecommand} (NAS message) and \textit{SecurityModeCommand} (RRC message), and similarly \textit{Securitymodecomplete} and \textit{SecurityModeComplete}. To ensure clear differentiation by the LLM, we explicitly renamed these messages to \textit{NAS\_SecurityModeCommand}, \textit{RRC\_SecurityModeCommand}, \textit{NAS\_SecurityMode\-Complete}, and \textit{RRC\_SecurityModeComplete}.

We then adopted a window-based approach with overlapping windows. Due to the absence of a clearly defined optimal window size (as identified by challenge C2), and the limitations of prior work, which considered only a fixed window size (size 6), we empirically evaluated window sizes ranging from 1 to 10. Specifically, the window size of 1 corresponds to the scenario without overlap where each message is processed independently without explicitly providing preceding messages in the context. Conversely, the window size of 10 indicates that each new incoming message is combined with the nine preceding messages for detection. We intentionally restrict each window to containing only one new incoming message at any given time. This decision facilitates streaming-based detection, ensuring immediate processing of incoming messages. Avoiding the incorporation of multiple new messages within a single window prevents the inherent latency of batch-processing approaches. These delays could potentially allow malicious activities to remain undetected for extended periods. Thus, our preprocessing strategy is explicitly designed to prioritise both responsiveness and effectiveness in anomaly detection.

\subsection{Phase 2: LLM Formatter and Message Extractor}
\begin{figure}
    \centering
    \includegraphics[width=1\linewidth]{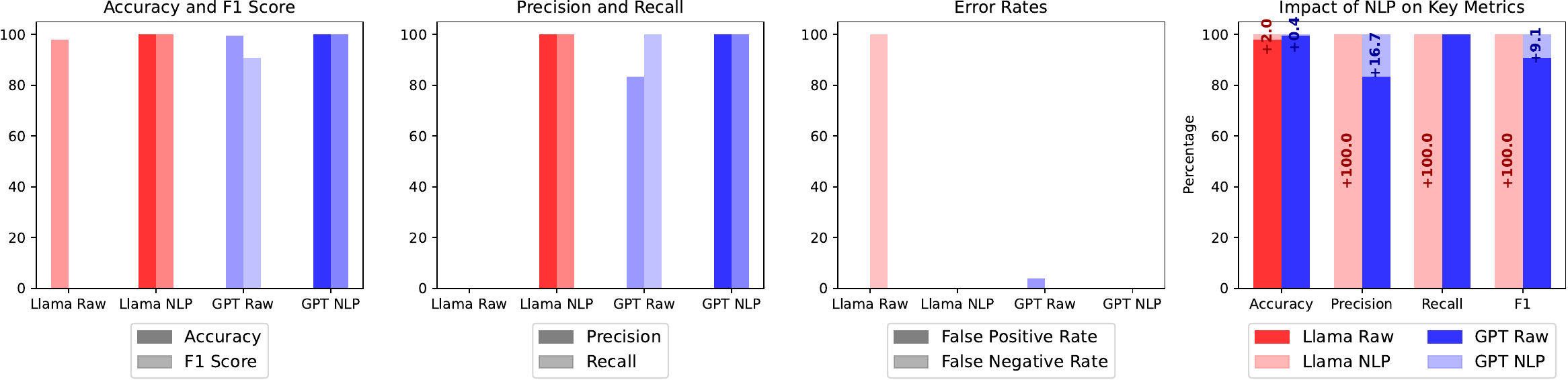}
\caption{Comparison of the impact of NLP friendliness (Raw data vs NLP-friendly formatted data) with Llama-3.1-8B-Instruct and GPT4o (Window size = 1)}
    \label{fig:raw_vs_nlp}
\end{figure}

Initial experimental results revealed that LLMs struggle to effectively perform anomaly detection tasks when provided with raw, structured data (as identified by challenge C4). Specifically, when raw data inputs such as \[RRCSetupRequest, 123456, 98765\] were directly fed into the LLM, the models failed to detect any attacks. To address this limitation, we introduced an LLM Formatter, which is designed to transform raw input data into a more LLM-friendly format. Instead of presenting the data as isolated fields, the formatter reconstructs them into coherent sentences, such as
\[
RRCSetupRequest\,\text{with RNTI 123456, TMSI 987654}
\]
Although this formatting adjustment appears minimal, it substantially improved the LLM’s performance. As depicted in Figure \ref{fig:raw_vs_nlp}, detection results (for window size 1) significantly increased after applying NLP-friendly formatting. Specifically, the LLM transitioned from completely failing to detect attacks with raw data inputs to successfully identifying all attacks (100\% Recall) when provided with formatted inputs.
This highlights the need for input formatting when leveraging LLMs for structured anomaly detection tasks.

Initial results showed that the Zero-shot approach failed to detect any attack accurately, which invalidate our assumption that an LLM will implicitly maintain state across a single session. Thus, we explicitly provide the previous message that shares the same TMSI as the current message, which significantly increases the detection performance as we discuss in Evaluation section.

\subsection{Phase 3: Detection Prompt Creator}
\label{sec:context}
Providing the correct prompt is the most critical step to achieving effective detection. We combined multiple components including the LLM-friendly formatted message window, previous message with the same TMSI value (if available), and the attack description to construct the detection prompt. Our initial experiments uncovered that even when running on the same LLM session, the LLM itself does not always maintain the full context as expected. Further, given that we have relaxed the ordering assumption of the message sequence, unlike other works, we explicitly provide the previous RRC/NAS message associated with the current message's TMSI value. We leverage a combination of system, user, and assistant prompts in integrating these components into the final detection prompt. In particular, the system prompt sets the scenario, expectations, attack description and desired formats, the user prompt provides the details (formatted records, and previous corresponding message), and the assistant prompt models desired response. This prompt provides sufficient contextual information for the LLM to differentiate between normal and attack scenarios. 

We construct the prompt explicitly to guide the LLM to classify the input data as either "Normal" or "Anomalous" (with an explanation). This explicit instruction enables the interpretability and accountability of the model's outputs. After construction, the prompt is sent to the LLM, which produces the classification (and explanation).

\subsection{Phase 4: Attack Extractor}
\label{sec:attack_extract}
The attack extractor generates the attack description that guides the zero-shot detection process. We support two complementary modes of operation for this phase. In the first mode, a human expert composes the attack description by summarising the main idea of the attack. This option allows domain specialists to encode their expert knowledge directly into the system.%, which relies on responsible person/team's technical knowledge.
In the second mode, we automate the extraction of attack descriptions using LLM-based agents. In particular, the LLM agent is provided with relevant information about the attack (e.g., textual reports, specifications, or documented vulnerabilities). The agent processes this input to extract the underlying idea of the attack and composes a concise description of the attack which can then be used for the detection. This automated approach reduces human effort and improves scalability compared to the first approach. This design can continuously accumulate new attack knowledge and generate usable descriptions.

\begin{lstlisting}[style=attackdesc, caption={Example attack descriptions},label={lst:attack-descriptions}]
% Manual crafted description
The adversary sends a RRCSetupRequest using a TMSI value of an existing connection and a new RNTI value. 

% LLM-generated description
The adversary assumes the victim's TMSI, sends a RRCSetupRequest to the base  station, and the base station, without integrity protection, deletes the victim's RRC security context due to the impersonation of the victim UE, thus disconnecting the victim from the network.
\end{lstlisting}

Listing~\ref{lst:attack-descriptions} depicts a manually crafted and LLM generated attack descriptions for the Blind DoS attack. This flexibility of the attack description generation can be leveraged to balance the accuracy (through expert-authored descriptions) with scalability (through LLM-based agents), which makes the attack extraction phase practical for deployment in dynamic 5G/O-RAN environments.

\subsection{Phase 5: Detector}
The detector is the component responsible for executing the classification step based on the prompt created in Phase 3. The detection prompt, which composes of the formatted message record, the retrieved previous message associated with the same TMSI, and the attack description, is inputted to the LLM. The LLM processes this prompt and produces the classification outcome as either ``Normal'' or ``Anomalous''. In the classification is ``Anomalous'', the LLM also generates a concise explanation. This explanation improves interpretability and allows operators to validate why a given message window was classified as malicious.

The classification result can then be leveraged in downstream O-RAN control context such as other xApps that can automatically trigger appropriate mitigation actions, such as updating access policies, isolating the malicious session, or adjusting scheduling to minimise the impact of the attack. Our design ensures both actionable anomaly detection and a seamless integration with existing O-RAN workflows.

\section{Implementation}
\subsection{Dataset}
Currently, there is no publicly available, real-world, large-scale 5G Layer-3 dataset, posing a significant challenge (C3) for the evaluation of Blind DoS detection solutions in practical scenarios. Therefore, to address this limitation, we utilise the dataset from previous work \cite{6gxsec}, which includes data collected from a physical testbed comprising four different commodity 5G smartphone models (Google Pixel 5 and 6, Samsung Galaxy A22, and Samsung Galaxy A53). Additionally, this dataset integrates data generated using the COLOSSEUM~\cite{colosseum} wireless network emulator.

However, the original dataset contains only three occurrences of Blind DoS attacks, which limits comprehensive evaluation. Thus, we augmented the dataset by adding 17 additional Blind DoS attack occurrences, bringing the total to about 1\% of attacks. More specifically, we added 17 \textit{RRCSetupRequest} messages with preselected existing TMSI and random RNTI values. Consequently, our enhanced dataset consists of 1,016 RRC and NAS messages including 20 Blind DoS attack scenarios, which enables more extensive and realistic security analyses of 5G Layer-3 communications. The original dataset was structured in an ordered fashion, where it group messages sequentially by each UE (e.g., $msg_1^{UE_1}, msg_2^{UE_1}, \ldots, msg_1^{UE_2}, msg_2^{UE_2}, \ldots, msg_1^{UE_3}, msg_2^{UE_3}, \ldots$). To reflect more realistic network conditions, we shuffled the dataset while preserving the message order for each individual UE (i.e., $msg_1^{UE_1}, msg_1^{UE_2}, msg_2^{UE_2}, msg_1^{UE_3}, \ldots$).

In our evaluation methodology, when the window size exceeds one message, we consider a window to be attacked if at least one message within that window is an attack message. For example, considering a window size of two messages, if one of the messages is identified as a Blind DoS attack, the entire window is labelled as attacked, regardless of the other message's status. Table~\ref{tab:window-stats} summarises the number of attack windows corresponding to different window sizes, clearly demonstrating how the number of attack windows increases with larger window sizes. However, we denote that, in practice, detecting the first attack window (out of all consecutive windows containing the Blind DoS) is the most appropriate, as it prevents the impact, whereas later detections are too late for effective prevention.

\subsection{LLM}
\label{sec:llms}
We evaluated several state-of-the-art open-source and commercial LLMs, including Llama-3.1-8B-Instruct, Microsoft Phi-3.5-mini, MistralAI Mistral-8B-2410, OpenAI GPT-4o, and deepseek-ai/DeepSeek-R1-Distill-Llama-8B. The open-source models were obtained from HuggingFace\cite{huggingface} and deployed using vLLM\cite{vLLM} on an NVIDIA A100 GPU (80GB). We utilised the official OpenAI API, for GPT-4o. The temperature parameter was set to 0 across all evaluations to ensure deterministic outputs and prevent LLMs from hallucinations or creative deviations in the predictions.

However, initial experimental results revealed significant performance differences among these models. Specifically, Phi-4-mini, Mistral-7B, and DeepSeek-R1-Distill-Llama-8B exhibited significantly inferior detection performance compared to Llama-3.1-8B-Instruct and GPT-4o. Thus, in the evaluation section, we primarily focus on presenting detailed results obtained from the Llama-3.1-8B-Instruct model. Further discussion regarding the practicality and deployment considerations of all evaluated models is provided in Section~\ref{sec:practicality}.

\section{Evaluation}
To evaluate the performance of the proposed framework, we aim to answer the following research questions. 
\begin{enumerate}[label=\textbf{RQ-\arabic*}, leftmargin=*, labelwidth=3em, labelsep=1em, align=left]
    \item Can a Zero-shot approach with existing LLMs effectively detect Blind DoS attacks? 
    \item Does having more context and combining Zero-shot with CoT improve the detection capability of LLM-based approaches? 
    \item How sensitive is detection to the semantics of the attack description?
    \item How does the effectiveness of LLM-based Blind DoS detection compare to traditional ML-based approaches? 
    \item How practical is it to leverage LLM-based methods for Blind DoS attack detection within the O-RAN context? 
\end{enumerate}

We primarily use Accuracy, Precision, Recall, F1 score, False Positive Rate (FPR), and False Negative Rate (FNR) for the evaluation. However, we note that achieving a high F1 Score along with low FPR and FNR is the desired outcome.

\subsection{Effectiveness of Zero-shot Approach}
\label{sec:rq1}
We first evaluated whether existing LLMs could detect Blind DoS attacks using a Zero-shot approach at window size $w{=}1$ without explicit training or examples. We compared two configurations: (i) \emph{without} explicitly providing the previous message (relying on the same LLM session to implicitly track context and state), and (ii) \emph{with} the previous message that shares the same TMSI as the current record.

Our results showed that without the previous message (Table \ref{tab:rq1-zeroshot}), the Zero-shot approach failed to detect any attack accurately. Despite the high accuracy of $98.03\%$ (due to imbalance of the dataset), the model produced $0\%$ precision, $0\%$ recall, $0\%$ F1, $0\%$ FPR, and $100\%$ FNR. In contrast, when we explicitly included the most recent message with the same TMSI, the detector achieved perfect detection performance of $100\%$ F1.

\begin{table}[!tbh]
\centering
\small
\caption{Blind DoS detection with/without the prev message}
\begin{tabular}{lcccccc}
\toprule
\textbf{Setting} & \textbf{Acc} & \textbf{Prec} & \textbf{Rec} & \textbf{F1} & \textbf{FPR} & \textbf{FNR} \\
\midrule
Without prev msg & 0.98 & 0 & 0 & 0 & 0 & 1 \\
With prev msg    & 1 & 1 & 1 & 1 & 0 & 0 \\
\bottomrule
\end{tabular}

\label{tab:rq1-zeroshot}
\end{table}

To assess model sensitivity in the zero-shot setting, we evaluated additional LLMs (comparable number of parameters) at window size=1 using the same approach. Results in Table~\ref{tab:rq1-w1-models} show clear variability across models. Specifically, all models achieved 100\% recall. However, FPR varied significantly resulting inferior F1 values. 

\begin{table}[!tbh]
\centering
\small
\caption{Blind DoS detection with different LLMs}
\begin{tabular}{lccccccc}
\toprule
\textbf{Model} & \textbf{Acc} & \textbf{Prec} & \textbf{Rec} & \textbf{F1} & \textbf{FPR} & \textbf{FNR} \\
\midrule
Llama3.1-8B & 1 & 1 & 1 & 1 & 0 & 0 \\
Qwen2.5-7B & 0.97 & 0.4 & 1 & 0.57 & 0.03 & 0 \\
DeepSeek-R1 & 0.33 & 0.03 & 1 & 0.06 & 0.67 & 0 \\
Phi-4-mini & 0.77 & 0.08 & 1 & 0.15 & 0.24 & 0 \\
Mistral-7B & 0.74 & 0.07 & 1 & 0.13 & 0.27 & 0 \\
\bottomrule
\end{tabular}
\label{tab:rq1-w1-models}
\end{table}

Our results indicate that a Zero-shot approach with existing LLMs can effectively detect Blind DoS attacks. However, the LLM must be provided with sufficient and explicit contextual information. The assumption that an LLM will implicitly maintain state across a session for complex, domain-specific tasks like anomaly detection in telecommunications is incorrect. Thus, the detection prompt must be carefully designed to include all relevant details, which allows the LLM to make informed classifications. Further, detection results also depend on the choice of LLM despite their similar model complexity.

\subsection{Effectiveness of Larger Context and Combining with CoT}
Intuitively, providing LLMs with additional context (historical message sequences) should facilitate more informed decision-making, improving their ability to detect Blind DoS attacks. To evaluate this hypothesis, we assessed the model's detection performance using various window sizes (1 to 10), where larger window sizes indicate more historical context provided to the model.

\begin{figure}
    \centering
    \includegraphics[width=1\linewidth]{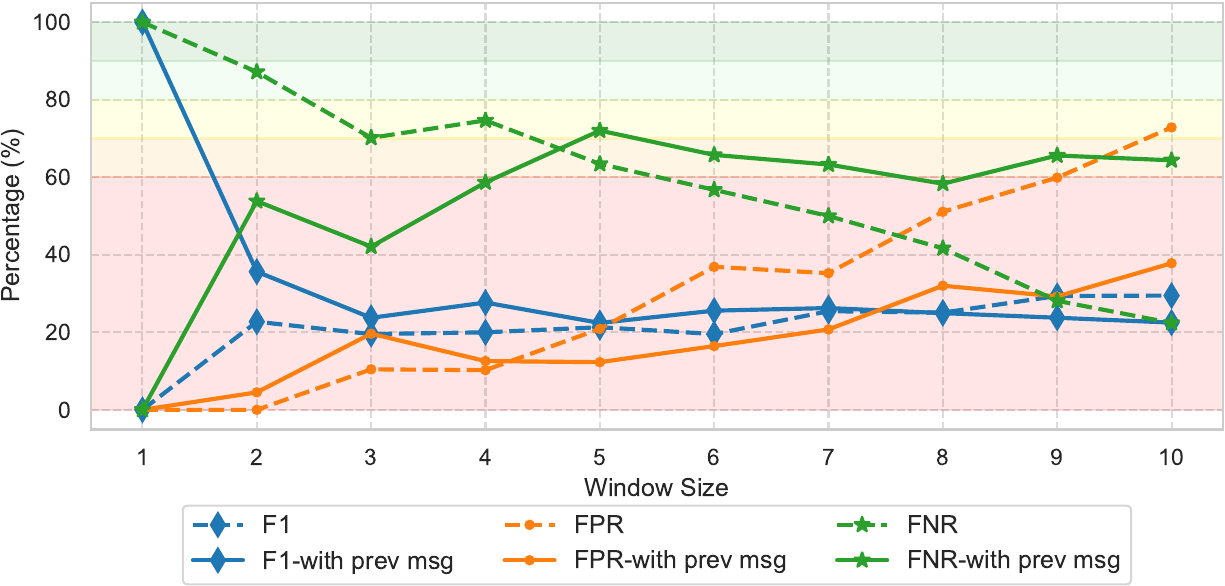}
    \caption{Detection performance with Zero-shot approach with and without previous message over different window sizes (Llama-3.1-8B-Instruct)}
    \label{fig:zeroshot}
\end{figure}

We first established a zero-shot baseline without any CoT prompting. In particular, we tested two configurations. First without explicitly providing the previous message (relying on implicit session memory), and then with the previous message sharing the same TMSI as the current message. Without the previous message, detection performance declines as window size increases, as shown in Figure~\ref{fig:zeroshot}. At window size 1, accuracy is high (98.03\%). However, precision, recall, and F1 are 0\% due to the 100\% FNR. Recall improved with the increased window size (e.g., 58\% at window size 8, 78\% at window size 10). Nevertheless, FPR significantly increased as well from 0\% at window size 1 to 73\% at window size 10. The maximum F1 was 29.5\% at window sizes 9 and 10.
With the previous message, window size 1 produced optimal detection performance, which resulted 100\% F1. However, increasing the window size degraded performance where F1 dropped to 35.6\% (FPR=4.5\%) when window size was 2. F1 further reduced (e.g., 23.7\% at window size 3, 22.5\% at window size 10) while FPR increased (up to 37.8\% at window size 10) with the further increase in window size. 

\begin{figure}
    \centering
    \includegraphics[width=1\linewidth]{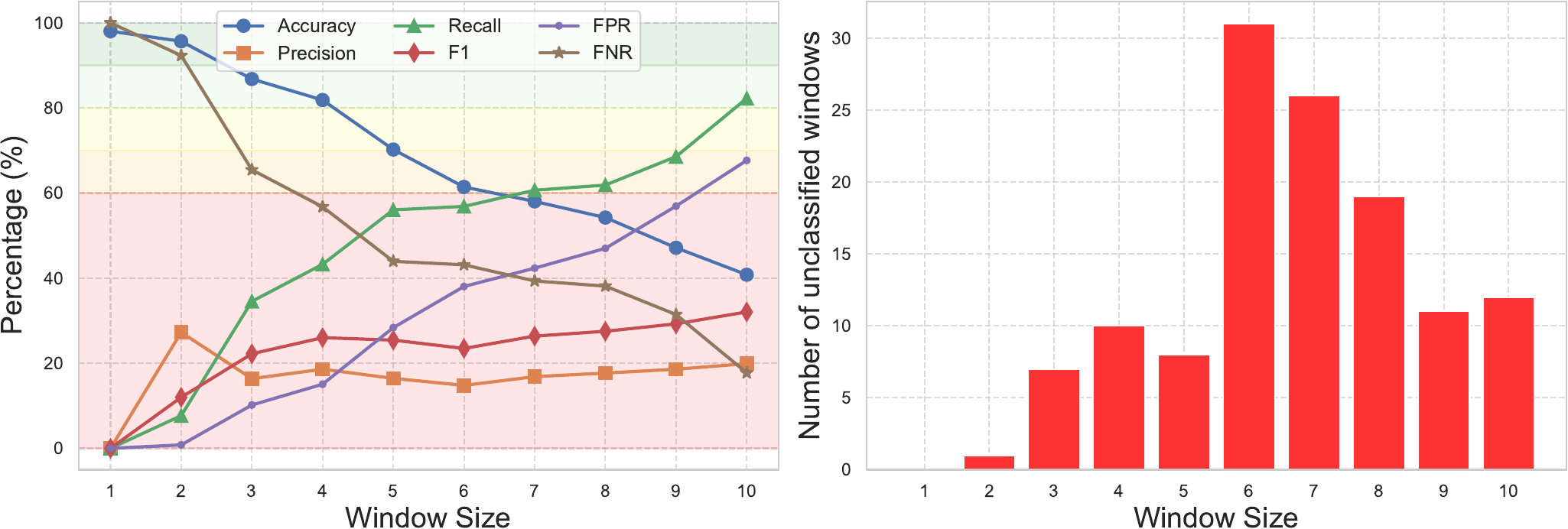}
    \caption{Detection performance with Zero-shot + Simple CoT approach (Llama-3.1-8B-Instruct)}
    \label{fig:zero_cot}
\end{figure}

Then we evaluated the effectiveness of combining the Zero-shot approach with generic CoT reasoning, given that recent studies have shown that LLMs improve Zero-shot reasoning tasks when simple CoT prompts are added, such as ``\textit{Let's think step by step}" before each answer~\cite{kojima2022large}. As shown in the left subplot of Figure~\ref{fig:zero_cot}, generic CoT did not consistently enhanced detection. Recall increased with window size (from ~0\% at window size 1 to 82\% at window size 10). However, FPR also increased significantly. The maximum F1 was only 32.1\% at window size 10. 

We have also observed that the model failed to classify some input windows (either as Normal or Anomalous) with the generic Zero-shot CoT approach. As shown in the right subplot of Figure~\ref{fig:zero_cot}, a significant number of windows remained unclassified. More specifically, at a window size of 6, the model failed to classify approximately 30 windows, which is about 2\% of the total data and 26 and 19 at windows size 7 and 8, respectively. Even though the number of unclassified windows were comparatively low in other window sizes, the count remained over 5 across other window sizes. The model either asked for more information in these instances or generate inconsistent outputs. Having a large number of unclassified windows hinders the practicality and usability of the generic Zero-shot CoT approach for reliable Blind DoS attack detection. Thus, we conclude that the overall performance of the generic Zero-shot CoT approach remains insufficiently robust or reliable for practical deployment.

\begin{figure}
    \centering
    \includegraphics[width=1\linewidth]{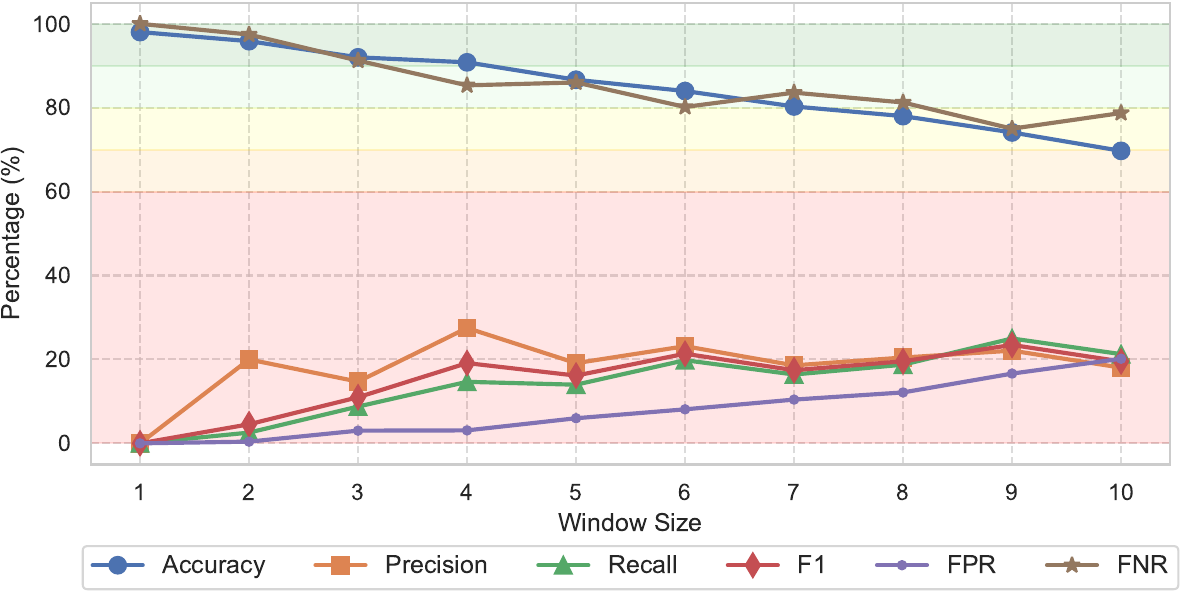}
    \caption{Detection performance with Zero-shot + Advanced CoT approach (Llama-3.1-8B-Instruct)}
    \label{fig:pure_cot}
\end{figure}

Next, we evaluated the impact of employing a customised CoT approach, explicitly designed for Blind DoS attack detection (specific prompts used are shown in the Appendix). Our intuition was that using problem-specific reasoning would guide the LLM more effectively, improving detection accuracy and overall model reliability compared to generic CoT approach. Nevertheless, experimental results shown in Figure~\ref{fig:pure_cot} did not support this. Although the customised CoT method fully eliminated the issue of unclassified windows,% (with zero unclassified windows across all window sizes), 
it reduced detection performance. Specifically, recall increased with larger window sizes, just 25\% at window size 9. Precision remained low, approximately between 18\%–23\%. Further, FPR exceeds 10\% for window sizes larger than 7 and reached to about 20\% at window size 10. The maximum F1 was 23.5\% at window size 9. Thus, despite customised CoT reasoning mitigated some issues (reducing FPR and unclassified windows), it still failed to address the primary goal of effective Blind DoS attack detection with larger context.

These results contradict the initial intuition on improvements by providing more context (larger window sizes) or using CoT prompting.% did not improve Blind DoS detection. 
Instead, these approaches reduced precision by increasing FPR and, for generic CoT, introduce unclassified outputs. Thus, a targeted, minimal context (with the previous relevant message) at window size 1 is most effective. 

\subsection{Detection Sensitivity vs Attack Description}
We next evaluated whether the wording and semantic of the attack description affect zero-shot detection performance. We extracted the original description of the Blind DoS attack from \cite{5GReasoner} and asked an LLM to generate concise descriptions of the attack. We repeated this procedure across six open-source LLMs (Llama-3.1-8B-Instruct, Llama-3.2-1B-Instruct, Llama-3.2-3B-Instruct, Qwen2.5-7B-Instruct, Mistral-7B-Instruct-v0.3, and Phi-4-mini-instruct) with temperature set to $1$ (standard level of randomness), producing $150$ Blind–DoS descriptions, 25 from each LLM. Each description is intended to capture, in natural language, the essential steps of the attack while remaining short enough to be used directly in the detection prompt.

To evaluate the detection performance against these attack descriptions, we grouped the $150$ descriptions by their coverage of five core predicates of the Blind DoS attack that we defined as below.
\begin{enumerate}[label=\textbf{P\arabic*}, leftmargin=*, labelwidth=3em, labelsep=1em, align=left, itemsep=0.01em]
\item Trigger message is explicitly identified (\texttt{RRCSetupRequest}).
    \item Mentions TMSI (or synonyms).
    \item States the TMSI is bound to a legitimate/victim UE (i.e., reuse of an \emph{in-use} TMSI).
    \item Explicitly indicates spoofing/impersonation/reuse of that TMSI.
    \item Mentions that the RNTI differs from the legitimate UE’s.
    \item Mentions the lack/absence of integrity protection for the relevant RRC message.
\end{enumerate}
Although we define \textbf{P5}, none of the generated descriptions satisfied it. Thus, we did not use \textbf{P5} in our alignment criteria below.

Using these, we defined four alignment groups as below.
\begin{itemize}[itemsep=0.1em]
    \item Directly aligned - P1 $\land$ P2 $\land$ P3 $\land$ P4
    \item Closely aligned - P1 $\land$ P2 $\land$ (P3 $\lor$ P4)
    \item Somewhat aligned - P1 $\land$ (P2 $\lor$ P3 $\lor$ P4 $\lor$ P6)
    \item Not aligned - Otherwise.
\end{itemize}
This resulted $37$ Directly aligned, $43$ Closely aligned, $14$ Somewhat aligned, and $56$ Not aligned descriptions.

For each of the $150$ descriptions (in four groups), we injected the text into the detection prompt in place of the attack description, while keeping all other components unchanged. We then ran the zero-shot detector over the dataset with window size of one and recorded the detection performance and compare distributions of F1 across groups. We also analyse the correlation between F1 and the number of predicates satisfied, and examine before/after results when missing predicates are completed by a simple linter. We present these groupings and results in the subsequent paragraphs.

As shown in Table \ref{tab:alignment_stats}, F1 increases monotonically with semantic alignment. Mean F1 (95\% bootstrap confidence interval) is 0.435, for Directly and 0.433 for Closely, Somewhat 0.144 for Somewhat, and 0.032 for Not align. Medians and $10^{th}$ percentiles show the same pattern. Further, only the top two groups resulted near-perfect runs. 

\begin{table}[htbp]
\centering
\caption{Summary Statistics of Groups}
\label{tab:alignment_stats}
\begin{tabular}{lrcccc}
\toprule
\textbf{Group} & \textbf{N} & \textbf{Mean} & \textbf{Median} & \textbf{$p_{10}$} & \textbf{Perfect} \\
\midrule
Directly & 37 & 0.435 & 0.333 & 0.229 & 2.7\% \\
Closely & 43 & 0.433 & 0.333 & 0.182 & 2.3\% \\
Somewhat & 14 & 0.144 & 0.095 & 0.000 & 0\% \\
Not & 56 & 0.032 & 0.000 & 0.000 & 0\% \\
\bottomrule
\end{tabular}
\end{table}

A Kruskal–Wallis test shows that the F1 distributions are different across the four semantic-alignment groups ($H{=}88.614$, $p{=}4.35{\times}10^{-19}$). Pairwise Mann–Whitney tests with Holm correction and Cliff’s $\delta$ effect sizes shows a step-wise semantic distribution where 
$Not < Somewhat < (Closely \approx Directly)$.
Differences between the Not align group and other groups are large where it was $\delta$=.937 for Direct, $\delta$=.946 for Closely, and $\delta$=.483 for Somewhat. Moreover, as depicted in Table \ref{tab:rq5-correlations}, F1 shows a strong monotone correlation with predicate coverage. In contrast, the description length is uncorrelated with F1.

\begin{table}[htbp]
\caption{Correlation comparison}
\centering
\begin{tabular}{lcc}
\toprule
\textbf{Correlation} & \textbf{Coefficient} & \textbf{$p$-value} \\
\midrule
Spearman $\rho$ (predicates, F1) & $0.719$ & $3.74\times10^{-25}$ \\
Kendall $\tau$ (predicates, F1)  & $0.578$ & $2.77\times10^{-18}$ \\
Spearman $\rho$ (length, F1)     & $-0.028$ & $0.737$ \\
\bottomrule
\end{tabular}
\label{tab:rq5-correlations}
\end{table}

These results indicate that the zero-shot based detection approach is rather semantics-sensitive and not phrase sensitive. High and stable detection performance could be achieved given the inclusion of appropriate predicates that clearly reflect the attack. On contrary, omitting the essential predicates significantly deteriorate detection performance.

\subsection{Compare with Traditional Approaches}
In this section we compare the effectiveness of LLM-based Blind DoS detection against the traditional ML-based approaches. In particular, we compare our proposed Zero-shot LLM detector against the AutoEncoder (AE) model adapted from 6G-XSec~\cite{6gxsec}, which is the best-performing traditional ML approach from recent literature. For a fair comparison, we evaluated both methods under the same unordered setting while varying the window size from 1 to 10. Further, we integrated the previous message with same TMSI as of the latest message into AE as a separate feature. We denote that each AE model was trained using approximately 2/3 of our dataset (700 messages) given that AE models require training phase. The entire dataset was utilised for evaluation. However, the LLM-based approach, did not require explicit training or threshold tuning.

Table~\ref{tab:comparison_AE_LLM} compares the detection performance of AE and LLM across all tested window sizes. The results clearly demonstrate that the LLM-based approach is highly effective with minimal context. At a window size of one, the LLM achieves perfect detection with an Accuracy, Precision, Recall, and F1 score of 1, and an FPR and FNR of 0. In contrast, the traditional AE-based method, could not achieve perfect scores, and only reached an F1 of 0.8 (Precision of 0.67, Recall of 1.00, and an FPR of about 1\%). However, as the window size increases, detection performance differs. More specifically, the LLM's detection deteriorate while the AE's detection improves. For example, at a window size of three, the AE produced an F1 score of 0.397, which is significantly higher than the LLM's F1 of 0.237. This trend continues, and AE records an F1 of 0.530, while the LLM drops to an F1 score of 0.225 at windows size 10.

Nevertheless, given its clear superiority at a minimal window size, the LLM-based approach provides a distinct advantage for low-latency O-RAN pipelines with minimal context and its operational attractiveness (no training, no thresholds, and robustness (discuss in next section)). In contrast, the AE consistently outperforms the LLM for larger windows. 

\begin{table*}
   \centering
    \caption{Comparison of AE and LLM detection performance}
    \begin{adjustbox}{max width=\linewidth}
    \begin{tabular}{c|cccccc|cccccc}
        \toprule
        \textbf{Window size} & \multicolumn{6}{c|}{\textbf{AE}} & \multicolumn{6}{c}{\textbf{LLM}} \\
        & \textbf{Accuracy} & \textbf{Precision} & \textbf{Recall} & \textbf{F1} & \textbf{FPR} & \textbf{FNR} & \textbf{Accuracy} & \textbf{Precision} & \textbf{Recall} & \textbf{F1} & \textbf{FPR} & \textbf{FNR} \\
        \midrule
        1  & 0.9902 & 0.6667 & \textit{\color{blue}{1.0000}} & 0.8000 & 0.0100 & \textit{\color{blue}{0.0000}} &
             \textit{\underline{\color{blue}{1.0000}}} & \textit{\underline{\color{blue}{1.0000}}} & \textit{\color{blue}{1.0000}} & \textit{\underline{\color{blue}{1.0000}}} & \textit{\underline{\color{blue}{0.0000}}} & \textit{\color{blue}{0.0000}} \\
        2  & 0.9182 & 0.2556 & \textit{\underline{\color{blue}{0.5897}}} & \textit{\underline{\color{blue}{0.3566}}} & 0.0686 & \textit{\underline{\color{blue}{0.4103}}} &
             \textit{\underline{\color{blue}{0.9360}}} & \textit{\underline{\color{blue}{0.2903}}} & 0.4615 & 0.3564 & \textit{\underline{\color{blue}{0.0451}}} & 0.5385 \\
        3  & \textit{\underline{\color{blue}{0.8471}}} & \textit{\underline{\color{blue}{0.2550}}} & \textit{\underline{\color{blue}{0.8947}}} & \textit{\underline{\color{blue}{0.3969}}} & \textit{\underline{\color{blue}{0.1557}}} & \textit{\underline{\color{blue}{0.1053}}} &
             0.7909 & 0.1493 & 0.5789 & 0.2374 & 0.1964 & 0.4211 \\
        4  & 0.7838 & \textit{\underline{\color{blue}{0.2447}}} & \textit{\underline{\color{blue}{0.9200}}} & \textit{\underline{\color{blue}{0.3866}}} & 0.2271 & \textit{\underline{\color{blue}{0.0800}}} &
             \textit{\underline{\color{blue}{0.8401}}} & 0.2081 & 0.4133 & 0.2768 & \textit{\underline{\color{blue}{0.1258}}} & 0.5867 \\
        5  & 0.7500 & \textit{\underline{\color{blue}{0.2633}}} & \textit{\underline{\color{blue}{0.9570}}} & \textit{\underline{\color{blue}{0.4130}}} & 0.2709 & \textit{\underline{\color{blue}{0.0430}}} &
             \textit{\underline{\color{blue}{0.8221}}} & 0.1871 & 0.2796 & 0.2241 & \textit{\underline{\color{blue}{0.1230}}} & 0.7204 \\
        6  & 0.7280 & \textit{\underline{\color{blue}{0.2842}}} & \textit{\underline{\color{blue}{0.9730}}} & \textit{\underline{\color{blue}{0.4399}}} & 0.3022 & \textit{\underline{\color{blue}{0.0270}}} &
             \textit{\underline{\color{blue}{0.7814}}} & 0.2043 & 0.3423 & 0.2559 & \textit{\underline{\color{blue}{0.1644}}} & 0.6577 \\
        7  & 0.7178 & \textit{\underline{\color{blue}{0.3090}}} & \textit{\underline{\color{blue}{0.9922}}} & \textit{\underline{\color{blue}{0.4712}}} & 0.3220 & \textit{\underline{\color{blue}{0.0078}}} &
             \textit{\underline{\color{blue}{0.7386}}} & 0.2043 & 0.3672 & 0.2626 & \textit{\underline{\color{blue}{0.2075}}} & 0.6328 \\
        8  & \textit{\underline{\color{blue}{0.7175}}} & \textit{\underline{\color{blue}{0.3293}}} & \textit{\underline{\color{blue}{0.9444}}} & \textit{\underline{\color{blue}{0.4883}}} & \textit{\color{blue}{0.3202}} & \textit{\underline{\color{blue}{0.0556}}} &
             0.6422 & 0.1780 & 0.4167 & 0.2495 & \textit{\color{blue}{0.3202}} & 0.5833 \\
        9  & \textit{\underline{\color{blue}{0.6974}}} & \textit{\underline{\color{blue}{0.3434}}} & \textit{\underline{\color{blue}{0.9938}}} & \textit{\underline{\color{blue}{0.5104}}} & 0.3585 & \textit{\underline{\color{blue}{0.0063}}} &
             0.6498 & 0.1815 & 0.3438 & 0.2376 & \textit{\underline{\color{blue}{0.2925}}} & 0.6563 \\
        10 & \textit{\underline{\color{blue}{0.6951}}} & \textit{\underline{\color{blue}{0.3612}}} & \textit{\underline{\color{blue}{0.9943}}} & \textit{\underline{\color{blue}{0.5299}}} & \textit{\underline{\color{blue}{0.3673}}} & \textit{\underline{\color{blue}{0.0057}}} &
             0.5760 & 0.1645 & 0.3563 & 0.2250 & 0.3782 & 0.6437 \\
        \bottomrule
    \end{tabular}
    \end{adjustbox}
    \label{tab:comparison_AE_LLM}
\end{table*}

\subsection{Practicality in O-RAN Context}
\label{sec:practicality}
Next we evaluate the practicality and feasibility of deploying LLM-based detection as an xApp running on the Near-RT RIC must satisfy timing constraints, where the response times must be less than 1 second (10ms - 1s). In particular, we measure the response time of Blind DoS detection using four locally deployed LLM models mentioned in Section \ref{sec:llms} with the zero-shot approach with Window size of one. We excluded GPT-4o from this analysis due to the fundamental difference in its deployment model (external API calls, which cause additional latency rather than just local inference).

\begin{figure}[!tbh]
    \centering
    \includegraphics[width=1\linewidth]{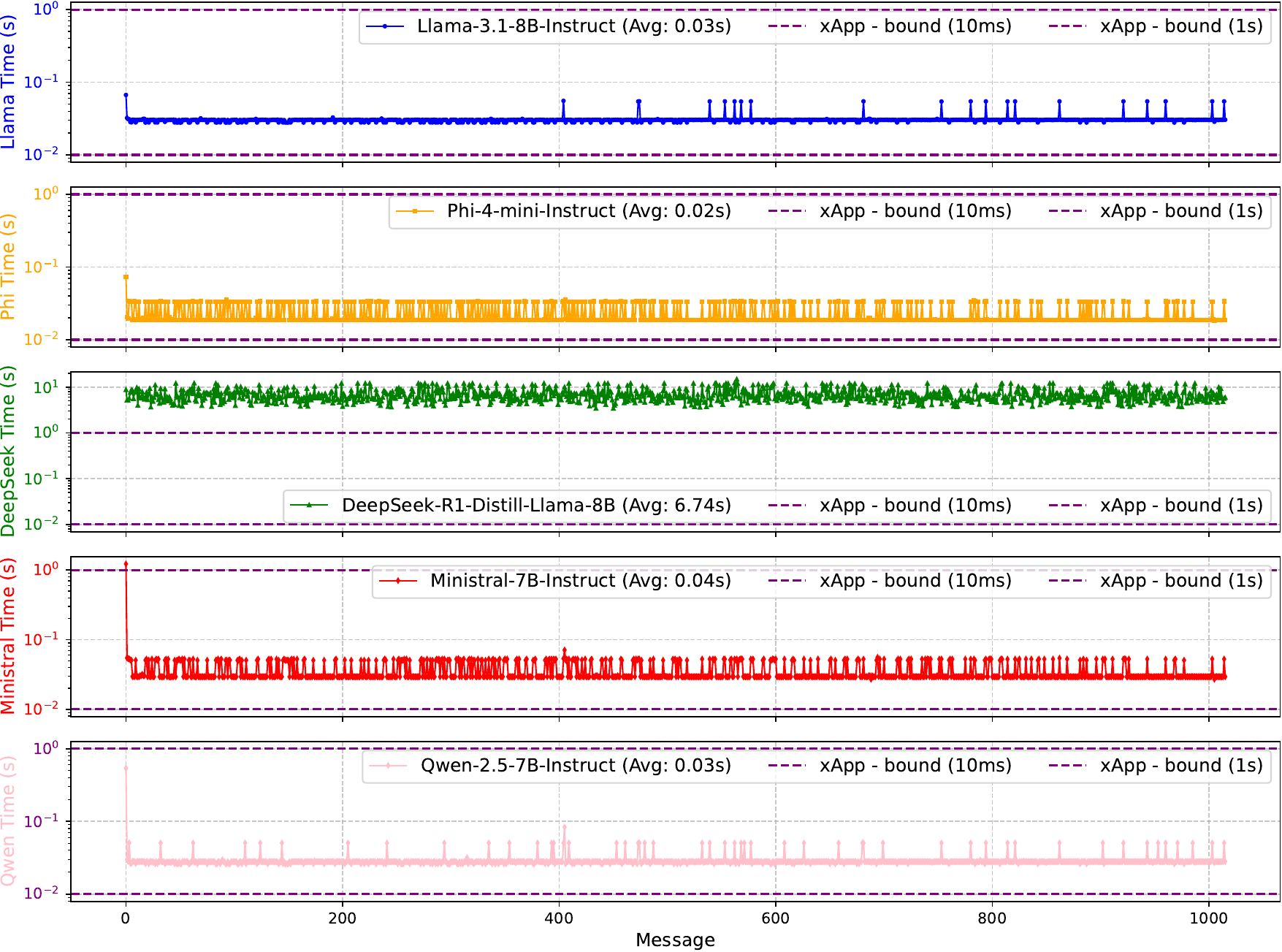}
    \caption{Comparison of LLM time consumption}
    \label{fig:timing}
\end{figure}

Figure~\ref{fig:timing} depicts the observed execution times (NLP processing + detection) across these models with the window size of one. The results show significant variations in performance across the tested models. DeepSeek-R1 constantly exceeds the one second time constraint required by near-RT RIC. DeepSeek-R1 shows an average detection time of 6.7 seconds. All other models average around 0.03 seconds. As shown in the figure, both Mistral and Phi models produced significant number of false positives (each spike reflects a message detected as an attack) despite their average detection times are well below the Near-RT RIC lower bound (10ms). Comparatively, Qwen produced less number of false positives. Llama-3.1-8B-Instruct produced the best detection with an average detection time of only 0.03s, which is well within the O-RAN timing requirements.

We have observed that each model exhibits higher processing times (spikes in the Figure) when analysing actual Blind DoS attack windows compared to normal traffic scenarios. This behaviour is caused by the higher number of tokens generated when LLMs classify a message as ``Anomalous'' rather than ``Normal''.

Considering additional processing overhead associated with data extraction, the overall detection process leveraging the Llama-3.1-8B-Instruct model remains practically feasible within the time constraints of Near-RT RIC xApps. Further improvements in inference speed can be achieved by leveraging advanced GPUs. 

\subsubsection{Robustness to Evasion Attacks}
We evaluate an evasion attack by altering five messages with Unicode confusables/full-width substitutions (2 Blind-DoS, 3 normal), simulating an adversary who manipulates the SDL (more details on Appendix). The AE model crashed at the feature-encoding step due to unencodable tokens, which is a known limitation of fixed-schema encoders~\cite{hypoglyph2025}. In contrast, the LLM-based approach continues to function on manipulated inputs without any preprocessing. Moreover, our LLM-based approach correctly classified all manipulated messages with F1 of 100\%. Given this robustness and operational effectiveness, LLM-based anomaly detection is strong alternative for traditional ML based approaches in O-RAN context.

\section{Discussion}
In this section, we discuss the limitations of our study and outline potential improvements and future work.

\subsection{Generalisability}
Although we initially designed the framework to detect Blind DoS attacks, it can be extended to detect additional Layer-3 attacks. To demonstrate this, we have also assessed its effectiveness against Null Cipher (Null)~\cite{5GReasoner}, Downlink DoS (DDoS) attacks~\cite{5GReasoner}, Downlink IMSI Extractor (DIMSI)~\cite{5GReasoner} under the same zero-shot setting (unordered data and window size $=1$). Table~\ref{tab:other_attacks} summarises results for three additional attacks using two LLMs.

\begin{table}[!htb]
\centering
\caption{Detection performance for other attacks}
\begin{tabular}{lccc}
\toprule
\textbf{Attack} & \textbf{Llama-3.1-8B-Instruct} (F1) & \textbf{GPT-4o} (F1)\\
\midrule
DIMSI  & $0$   & $1.0$\\
Null   & $1.0$ & $1.0$\\
DDoS   & $0$   & $0$\\
\bottomrule
\end{tabular}
\label{tab:other_attacks}
\end{table}

Attacks such as Null Cipher and Blind DoS, which reveal specific information from corresponding NAS/RRC messages, can be effectively detected using concise natural-language descriptions. However, DIMSI detection requires a more robust model (e.g., GPT-4o) and fails with Llama, indicating model sensitivity when descriptions are brief or implicit. Despite the influence of the LLM model choice, our LLM-based framework effectively detects attacks with consistent message-level cues.
Unlike Blind DoS, Null Cipher, or Downlink IMSI Extractor attacks, Downlink DoS lacks a single, identifiable attack message, as it involves overshadowing or dropping the response to a NAS \textit{AuthenticationRequest} message. Consequently, our current design, which relies on unordered data (current message and corresponding last message), cannot detect this attack. A potential solution is to introduce a short context window that captures recent messages in an ordered sequence.

\subsection{Enhanced Dynamic Prompt Construction} 
In first experiments, detection prompts were manually constructed from known attack signatures. However, this manual prompt construction approach reduces adaptability given the continuously evolving threat landscape. We therefore consider an automated dynamic prompt construction framework based on AI agents~\cite{zhou2024understanding,autogen,nvidia2025llmagents} that continuously generates natural-language attack descriptions which can be fed directly into our zero-shot detector. The original Blind DoS attack description's extract to the LLM was manually provided. However, current advances show that this process can be, at least partly, automated. For example, multiple AI agents can be leveraged to actively gather threat intelligence from the latest scholarly articles, cybersecurity bulletins, and news sources. A separate AI agent can then be utilised to analyse the collected threat intelligence and construct updated detection prompts continuously. Having an automated framework like that could
significantly enhance the adaptability of the detection mechanism.

Newly emerging (even zero-day) exploits can be potentially incorporated as soon as their mechanics are described in words, which avoids lengthy training or labelled datasets, given that our framework operates on natural-language descriptions rather than trained classifiers. However, to mitigate drift and hallucinations, we require to ensure the predicate coverage before deployment and fall back known-good prompt set when descriptions lack essential predicates, which can maintain the framework's adaptability and effectiveness.

\section{Related Work}
Several recent studies have focused on detecting attacks targeting cellular networks, particularly the Blind DoS attack. 5G-SPECTOR~\cite{5gsepctor} introduced a rule-based detection approach targeting Layer-3 attacks by constructing explicit detection rules. Despite the effectiveness of the rule-based methods, rule-based methods inherently require manual rule definitions and the continuous tracking of UE status/messages. 6G-XSec~\cite{6gxsec} adopted traditional ML models in identifying Blind DoS attacks and other Layer-3 attacks. They showed that Autoencoders, could achieve perfect detection performance. However, given the use of traditional ML models, this approach require training and inherently substantial dataset for the training. Scalingi et al. addressed Blind DoS detection using physical-layer features, emphasising signal-level patterns rather than protocol behaviours~\cite{detran}. Despite their effectiveness, these existing approaches share common limitations where rule-based methods require explicit rule definition and stateful tracking, whereas traditional ML models require significant training data and training. Moreover, these approaches lack inherent interpretability.

\section{Conclusion}
We present an LLM-based O-RAN compliant Blind DoS detection framework that operates in a zero-shot setting from natural-language attack descriptions. This framework can be used to overcome issues including the need for explicit rules, substantial amounts of data, and the training that exist with rule-based and traditional ML-based approaches and can be extended to generalise Layer-3 attacks in 5G and beyond. Further, the solution works with unordered data and is capable of providing clear explanations to increase the reliance on the detection approach and execute within the Near-RT RIC timing constraints. Moreover, we demonstrate that detection quality follows the semantic completeness of the description rather than verbosity or phrasing, and we proposed a dynamic prompt construction pipeline that automates the generation of attack descriptions. Additionally, we also highlight the practicality, limitations, and potential solutions considering the O-RAN context. Further empirical evaluations and modifications are needed for other types of attacks. Additionally, future work should assess detection performance when multiple attack descriptions are combined into a single prompt and techniques to improve the detection with larger window sizes.

\section*{Acknowledgement}
This research paper is conducted under the 6G Security Research and Development Project, as led by the Commonwealth Scientific and Industrial Research Organisation (CSIRO) through funding appropriated by the Australian Government’s Department of Home Affairs. This paper does not reflect any Australian Government policy position. For more information regarding this Project, please refer to \url{https://research.csiro.au/6gsecurity/}.

%-------------------------------------------------------------------------------
% optional clearing of the page
\cleardoublepage
\appendix
\cleardoublepage

\section*{Open Science}
\subsection*{Artefacts}
\href{https://github.com/thusithathilina/Desc2Det}{Github link} contains all the scripts that can be used to reproduce the detection result and the datasets that we used for evaluation. Further, we have also included the source code and Dockerfile to build the xApp. Also, this contains the instructions to deploy the xApp in OSC's Near-RT RIC platform.

\subsection*{Prompts}
Here we list exact prompts used throughout this work. The "Knowledge Cutoff Date" and "Today's Date" sections are implicitly generated by the vLLM framework, whereas other tags are explicitly generated by vLLM, given that we employed various prompt types (system, user, and assistant). All prompts listed below are based on the Llama-3.1-8B-Instruct model.

\subsubsection*{Zero-shot}
\begin{lstlisting}[basicstyle=\small,backgroundcolor=\color{gray!10}]
<|begin_of_text|><|start_header_id|>system<|end_header_id|>

Cutting Knowledge Date: December 2023
Today Date: 26 Jul 2024

You are an expert anomaly detecting assistant in the 5G context.
You know about the Blind DoS attack, where the adversary sends a RRCSetupRequest using a TMSI value of an existing connection and a new RNTI value.
You are given a set of messages between a gNB and multiple UEs in chronological order.
Based on the understanding of the given attack, you will need to determine whether the following message sequence contains a Blind DoS attack or not.
Remember, you must either say 'Normal' or 'Anomalous' without any explanation.<|eot_id|><|start_header_id|>user<|end_header_id|>

Previous message: RRCSetupRequest with RNTI 26168, and TMSI 0 (New message)<|eot_id|><|start_header_id|>user<|end_header_id|>

RRCSetup with RNTI 26168, and TMSI 0 (New message)<|eot_id|><|start_header_id|>assistant<|end_header_id|>
\end{lstlisting}

\subsubsection*{Zero-shot with Generic CoT}

\begin{lstlisting}[basicstyle=\small,backgroundcolor=\color{gray!10}]
<|begin_of_text|><|start_header_id|>system<|end_header_id|>

Cutting Knowledge Date: December 2023
Today Date: 26 Jul 2024

You are an expert anomaly detecting assistant in the 5G context.
You know about the Blind DoS attack, where the adversary sends a RRCSetupRequest using a TMSI value of an existing connection and a new RNTI value.
You are given a set of messages between a gNB and multiple UEs in chronological order.
Based on the understanding of the given attack, you will need to determine whether the following message sequence contains a Blind DoS attack or not.
Remember, you must either say 'Normal' or 'Anomalous' without any explanation.<|eot_id|><|start_header_id|>user<|end_header_id|>

Previous message: RRCSetupRequest with RNTI 26168, and TMSI 0 (New message)<|eot_id|><|start_header_id|>user<|end_header_id|>

RRCSetup with RNTI 26168, and TMSI 0 (New message)<|eot_id|><|start_header_id|>assistant<|end_header_id|>

Let's think step by step<|eot_id|><|start_header_id|>assistant<|end_header_id|>
\end{lstlisting}

\subsubsection*{Zero-shot with Custom CoT}
\label{sec:prompt_customcot}

\begin{lstlisting}[basicstyle=\small,backgroundcolor=\color{gray!10}]
<|begin_of_text|><|start_header_id|>system<|end_header_id|>

Cutting Knowledge Date: December 2023
Today Date: 26 Jul 2024

You are an expert anomaly detecting assistant in the 5G context.
You know about the Blind DoS attack, where the adversary sends a RRCSetupRequest using a TMSI value of an existing connection and a new RNTI value.
You are given a set of messages between a gNB and multiple UEs in chronological order.
Based on the understanding of the given attack, you will need to determine whether the following message sequence contains a Blind DoS attack or not.
Remember, you must either say 'Normal' or 'Anomalous' without any explanation.<|eot_id|><|start_header_id|>user<|end_header_id|>
Previous message: RRCSetupRequest with RNTI 26168, and TMSI 0 (New message)<|eot_id|><|start_header_id|>user<|end_header_id|>
RRCSetup with RNTI 26168, and TMSI 0 (New message)<|eot_id|><|start_header_id|>assistant<|end_header_id|>
Let's think step by step
Step 1: Check explicitly if an RRCSetupRequest message exists (mandatory for Blind DoS attack detection).
Step 2: Note the TMSI value used in this RRCSetupRequest.
Step 3: Check if a previous RRCSetupRequest (or subsequent message) exists with the same TMSI but a different RNTI.
Step 4: Classify the sequence explicitly as Anomalous if Step 3 condition is met and explain the reason. Otherwise classify as Normal.<|eot_id|><|start_header_id|>assistant<|end_header_id|>
\end{lstlisting}

\cleardoublepage
\bibliographystyle{plain}
\bibliography{references}
\cleardoublepage

\section{O-RAN Architecture}
\begin{figure}[htbp]
    \centering
    \includegraphics[width=\linewidth]{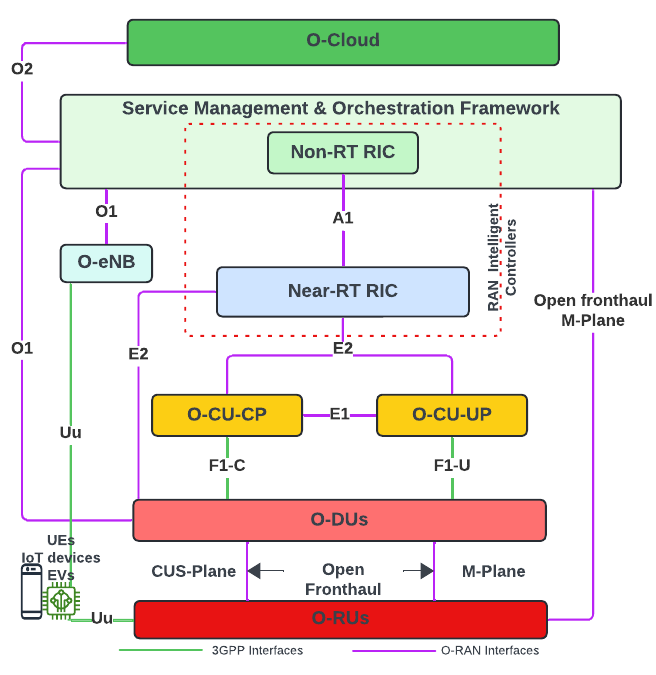}
    \caption{High-level O-RAN Logical Architecture \& Primary Interfaces}
    \label{fig:oran_arch}
\end{figure}

\section{Dataset}
\begin{table}[htbp]
\centering
\caption{Summary of attack windows by window size} 
\begin{tabular}{ccc}
\toprule
Window size & No. of windows & No. of attack windows\\
\midrule
1 & 1016 & 20 \\ 
2 & 1015 & 40 \\
3 & 1014 & 60 \\
4 & 1013 & 80 \\
5 & 1012 & 100 \\
6 & 1011 & 119 \\
7 & 1010 & 137 \\
8 & 1009 & 155 \\
9 & 1008 & 173 \\
10 & 1007 & 191\\
\bottomrule
\end{tabular}
\label{tab:window-stats}
\end{table}

\section{Explanations and Latency}
We enabled explanation generation and reran detection on the Blind DoS detection. All other configurations remain same where the detector receives unordered NAS/RRC messages (window size $=1$) and returns a classification except. However, instead of pure classification the LLM also generates a short natural–language rationale behind its decision. 
Similar to the standard scenario, across 1016 messages with 20 Blind DoS instances, we obtained  100\% F1. Nevertheless, generating explanations increases the number of generated tokens, which results an increase in classification time for anomalous instances as shown in Table \ref{tab:appendix-latency}. In particular, the mean classification time increased to 0.042s, where without explanation it was 0.031s. More specifically, the average classification time for 20 anomalies was increased to 0.7s from 0.05s. However, all the responses completed within 1s, which is the upper limit for Near-RT RIC operations. 

\begin{table}[t]
\centering
\caption{Latency with and without explanations}
\begin{tabular}{lrr}
\toprule
\textbf{Metric} & \textbf{With explanation} & \textbf{Without explanation} \\
\midrule
Mean (ms)          & 41.86  & 30.52  \\
Median (ms)        & 29.05  & 30.32  \\
P$_{90}$ (ms)      & 29.13  & 30.44  \\
P$_{95}$ (ms)      & 29.20  & 30.49  \\
P$_{99}$ (ms)      & 648.87 & 54.54  \\
Max (ms)           & 961.40 & 66.81  \\
Min (ms)           & 26.76  & 27.84  \\
$<\!1$ s (\%)       & 100.00 & 100.00 \\
\bottomrule
\end{tabular}
\label{tab:appendix-latency}
\end{table}

Below are some of the descriptions generated by the detector to justify its classification.
\begin{lstlisting}[basicstyle=\small,backgroundcolor=\color{gray!10},label=lst:nulldown,caption=Sample explanation generated by the franework,captionpos=b]
Anomalous. The RRCSetupRequest message with RNTI 63968 and TMSI 1777776 is attempting to reuse the TMSI from the 
previous RRC_SecurityModeComplete message, which is a characteristic of the Blind DoS attack.

Anomalous. The RRCSetupRequest message with RNTI 42397 and TMSI 72035 is using the TMSI from the previous RRC_SecurityModeComplete message, which is a characteristic of the Blind DoS attack.

Anomalous. The RRCSetupRequest message contains a TMSI value (59822) that matches the TMSI value of the previous RRC_SecurityModeComplete message, indicating an attempt to reuse an existing connection with a new RNTI value, which is a characteristic of a Blind DoS attack.
\end{lstlisting}
It seems that the LLM correctly learn/understand the attack from the provided natural-language attack description, according to the generated explanations. Explanation improves the interpretability of the model. However, given the increase in classification time, either we need to further limit the maximum number of output tokens or disable the explanation.

\section{Security and Privacy Considerations in O-RAN Context} 
Security and privacy concerns are essential factors to consider given the use of LLM and O-RAN context. In particular, we leverage subscriber identifiers, including TMSI and RNTI, which can result in potential privacy concerns if unauthorised entities gain access as these can be use to derive subscriber behaviours. Further, the use of a SDL can introduces additional security risks. For example, a malicious operator or user with access to the SDL (direct or through an xApp) could intentionally alter messages~\cite{dayaratne2024exploiting} to prevent detection or introduce errors in LLM-based classification by using different Unicode characters~\cite{nlpattack}. Such attacks could potentially bypass the LLM-based detection.

Therefore, future work should explore different types of attacks and their impact on LLM-based anomaly detection methods, considering the O-RAN context and its inherent vulnerabilities. Further, potential solutions, including adversarial training (use of malicious message windows and corresponding feedback), or smart access control to secure the SDL and its data, that can be used to mitigate the impact of those attacks require more investigations.

\subsection{Evasion Attacks on SDL}
Here we considered that the attacker's intention is to bypass existing ML detectors. However, instead of deleting or corrupting data, we assumed that an adversary leverage a malicious xApp to performs subtle Unicode-level substitutions (``hypoglyph'') to valid Layer-3 messages. For example, an adversary could modify ``RRCSetupRequest'' in a way it appears unchanged to a human, while internally replacing the Latin ``C'' (U+0043) with Cyrillic ``C'' (U+0421), the Latin ``e'' (U+0065) with Cyrillic ``e'' (U+0435), and the Latin ``q'' (U+0071) with a visually confusable character (e.g., U+055B).

%%%%%%%%%%%%%%%%%%%%%%%%%%%%%%%%%%%%%%%%%%%%%%%%%%%%%%%%%%%%%%%%%%%%%%%%%%%%%%%%
\end{document}